\documentclass[a4paper]{article}
\newenvironment{IEEEproof}{\begin{proof}}{\end{proof}}
\newcommand \techreport[1]{#1}
\newcommand \ieeetrans[1]{}

\makeatletter
\usepackage{paralist}
\usepackage{cite}
\usepackage[cmex10]{amsmath}
\usepackage{amsfonts}
\usepackage{amssymb}
\usepackage{amsbsy}
\usepackage{amsthm}
\usepackage{mathrsfs}
\usepackage{graphicx}
\usepackage{algorithm}
\usepackage{algorithmicx}
\usepackage{algpseudocode}
\usepackage{psfrag}
\usepackage{dsfont}
\usepackage{subfigure}

\newcommand \defn {\mathrel{\triangleq}}
\newcommand \ELb {\mathcal{L}}
\renewcommand\Pr{\mathop{\mbox{\ensuremath{\mathbb{P}}}}\nolimits}
\newcommand\E{\mathop{\mbox{\ensuremath{\mathbb{E}}}}\nolimits}

\newcommand\argmin{\mathop{\mathrm{argmin}}}
\newcommand \thr {\tau}
\newcommand \hthr {\hat{\thr}_n^*}
\newcommand \hns {\hat{n}^*}
\newcommand \ns {n^*}

\newcommand \err {\varepsilon}
\newcommand \noise {\omega}
\newcommand \loss {L}
\newcommand \worst {\mathbb{L}}
\newcommand \pa {p_A}
\newcommand \pu {p_U}
\newcommand \LA {\ell_A}

\newcommand \LU {\ell_U}

\newcommand \LB {\ell_B}
\newcommand \CX {\mathcal{X}}
\newcommand \decfun {q}
\newcommand \decs {g}
\newcommand \Decs {G}
\newcommand\ind[1]{\mathop{\mbox{\ensuremath{\mathbb{I}}}}\left\{#1\right\}}
\newcommand\set[1] {\left\{#1\right\}}
\newcommand\cset[2] {\left\{#1 \mathrel{:} #2\right\}}

\newcommand \auth {\mathcal{P}}
\newcommand \ver {\mathcal{V}}

\newcommand \id {I}
\newcommand \code {\Phi}
\newcommand \metric {\gamma}
\newcommand \minmetric {\gamma_{\min}}
\newcommand \Borel[1] {\mathcal{F}_{#1}}

\newcommand \rx {\hat{x}}

\newcommand\Reals {{\mathds{R}}}

\newcommand \Swiss {\textsc{Swiss-Knife}}
\newcommand \Hitomi {\textsc{Hitomi}}
\newcommand \ECMAD {\textsc{ECMAD}}
\newcommand \MAD {\textsc{MAD}}
\newcommand \Hancke {\textsc{HaKu}}

\theoremstyle{plain} \newtheorem{remark}{Remark}
\theoremstyle{plain} \newtheorem{definition}{Definition}
\theoremstyle{plain} 
\theoremstyle{plain} 
\theoremstyle{plain} 
\theoremstyle{plain} \newtheorem{theorem}{Theorem}
\theoremstyle{plain} 
\theoremstyle{plain} \newtheorem{lemma}{Lemma}
\theoremstyle{plain}

\title{Expected loss analysis of thresholded authentication protocols in noisy conditions}
\ieeetrans{
\author{Christos~Dimitrakakis,~\IEEEmembership{Member,~IEEE,}
  Aikaterini~Mitrokotsa
  and~Serge~Vaudenay,%
\thanks{C. Dimitrakakis is with FIAS in Frankfurt, Germany}%
\thanks{A. Mitrokotsa and S. Vaudenay are with EPFL in Lausanne, Switzerland.}
}
}
\techreport{
  \author{Christos Dimitrakakis \and Aikaterini Mitrokotsa \and Serge Vaudenay}
}

\begin{document}
\maketitle
\begin{abstract}
  A number of authentication protocols have been proposed recently,
  where at least some part of the authentication is performed during a
  phase, lasting $n$ rounds, with no error correction. This requires
  assigning an acceptable threshold for the number of detected
  errors. This paper describes a framework enabling an
  \textit{expected loss} analysis for all the protocols in this
  family. Furthermore, computationally simple methods to obtain nearly
  optimal value of the threshold, as well as for the number of rounds
  is suggested. Finally, a method to adaptively select both the number
  of rounds and the threshold is proposed.
\end{abstract}

\section{Introduction}

Traditionally~\cite{stallings:cryptography,stinson:cryptography},
authentication is assumed to be taking place on an error-free channel,
and error analysis is performed separately from cryptographic analysis
of protocols.  However, a number of authentication protocols have been
proposed~\cite{brands94, bussard, singelee1,tippenhauer,
  sheddingLight,
  cryptoeprint:2009:310,hancke05,reid2007,KimAKSP-2008-icisc}, where
at least some part of the authentication is performed during a
challenge-response phase lasting $n$ rounds with no error correction,
due to a need to detect relay attacks by timing delays.  The noise
necessitates the use of a tolerance threshold $\thr$, such that a
party is authenticated if the total error of its responses $\epsilon$
is below the threshold $\thr$.

This paper introduces a general framework for analysing such
protocols.  We assign a cost $\LA$ to the event that we authenticate a
malicious party $A$---which we call the {\em attacker}---a cost (or
loss) $\LU$ to the event that we fail to authenticate a valid party
$U$---which we call the {\em user}---and a cost $\LB$ for each round
of the challenge-response phase.  Our goal is to select $n, \tau$ so
as to minimise the \textit{expected loss} $\E L$.

The paper is organised as follows. Section~\ref{sec:relatedwork}
presents related work, while Section~\ref{sec:preliminaries}
introduces notation and thresholded authentication
protocols. Section~\ref{sec:loss-analysis} contains the
\textit{expected loss} analysis under noise. In particular,
Sec.~\ref{sec:threshold-choice} suggests a method to calculate the
threshold accompanied by a finite sample loss bound, while
Sec.~\ref{sec:rounds-choice} provides a further bound by selecting an
appropriate number of rounds. These results only require that the
expected error of the attacker is higher than that of the
user. Section~\ref{sec:specific-protocols} applies the above analysis
to a number of currently used protocols.  Section~\ref{sec:noise}
suggests a high-probability method for estimating the channel noise
and presents the results of simulation experiments that compare our
choice of threshold with thresholds derived using asymptotic
approximations. Finally, Sec.~\ref{sec:conclusion} concludes the
paper. For completeness, the appendix provides some useful auxilliary
results regarding the finite sample and the asymptotic derivations.

\subsection{Related Work}
\label{sec:relatedwork}
In certain authentication protocols part of the communication is
performed in noisy channels without employing error correction.
Specifically, a {\em rapid-bit exchange phase} was introduced
in~\cite{brands94} to compute an upper bound on the distance of the
prover $\auth$.  This is composed of $n$ challenge-response
\textit{rounds}, used to calculate a round-trip time and thus place a
bound on the distance.  Subsequently, a broad range of
\textit{distance bounding} protocols were proposed, both for
RFID~\cite{hancke05, KimAKSP-2008-icisc, reid2007, singelee1,
  sheddingLight}, as well as other wireless devices~\cite{capkun,
  capkun2, tippenhauer}.

Hancke and Kuhn \cite{hancke05} were the first to indicate that since
the \textit{rapid-bit exchange} phase is taking place in a noisy
channel, challenges and responses may be corrupted.  Thus, a
legitimate user may fail to get authenticated.  Their protocol (henceforth
\Hancke), employed $n$ rounds and authenticated any prover who made a
number of mistakes $\err$ less than an acceptance threshold $\tau$, so
as to reduce the number of false rejections.  Using the binomial
distribution and an assumption on the error rates they give
expressions for the \textit{false accept} and \textit{false reject
  probability} as a function of $n$ and $\tau$, but they provide no
further analysis.  Nevertheless, they indicate that the number of
challenge-response rounds $n$ in the rapid bit exchange phase should
be chosen according to the expected error rate.  Kim et
al. \cite{KimAKSP-2008-icisc} extend this approach with the {\Swiss}
protocol by considering {\em three} types of errors. 
Finally~\cite{singelee1}, rather than using a threshold $\tau$,
proposed a protocol (henceforth {\ECMAD}) using an error correcting
code (ECC).  {\ECMAD}, which extends the {\MAD}
protocol~\cite{capkun}, uses only $k$ of the $n$ total rounds for the
challenges and responses. The remaining $n-k$ rounds are used to
transmit the $(n,k)$ ECC.  This has the effect of achieving better
security (in terms of false acceptance rates) with the same number of
rounds $n$.

All these approaches use $n$ rounds in the noisy authentication phase.
However, they do not define the optimal $n$.  They simply state that
the probability of authenticating a user becomes much higher than the
probability of authenticating an attacker as $n$ increases.  However,
a large value of $n$ is incompatible with the requirements of many
applications and devices (i.e. high value of $n$ leads to high
overhead for resource-constrained devices). This can be modelled by
assigning an {\em explicit cost} to every round, which should take
into account the transmission energy, computation and time overhead.
This cost has so far not been explicitly taken into consideration.

Another work that is closely related to ours is~\cite{singelee2},
which, given a {\em required} false acceptance and false rejection
rate, provides a {\em lower bound} on the number of required
rounds. This analysis is performed for both {\Hancke} and
{\ECMAD}. However, it assumes that the number of rounds $n$ would be
large enough for the binomial distribution of errors to be
approximately normal.  Our analysis is more general, since it uses
finite-sample bounds that hold for any bounded error function.

Recently, Baign\`{e}res et al. \cite{ProvSec2010} have given an
analysis on the related topic of distinguishing between a real and a
fake solver of challenge-response puzzles. More precisely, they study
CAPTCHA-like protocols and provide a threshold which minimizes the
probability of error in these protocols.  The main differences between
the analysis presented in this paper and~\cite{ProvSec2010} can be
summarised below: \begin{inparaenum}[(a)]
\item We perform an expected loss analysis rather than an error
  analysis.
\item Our bounds hold uniformly, while~\cite{ProvSec2010} uses an asymptotically optimal distinguisher.
\item We consider bounded errors rather than $\{0,1\}$ errors for each
  challenge-response.
\item We additionally propose a method to estimate channel noise.
  This is of course not applicable in the context
  of~\cite{ProvSec2010}, due to the different setting.
\end{inparaenum}

A more general work on authentication under noisy conditions was
presented in~\cite{lai}.  This provided tight information-theoretic
upper and lower bounds on the attacker's success in impersonation and
substitution attacks, proving that it decreased with noise.  However,
our analysis shows that, when one considers losses due to
communication overhead and false rejections of users, the expected
loss increases, which is a natural result.

\subsection{Our contribution}
In this paper, we perform a detailed \textit{expected loss} analysis
for a general class of multi-round authentication protocols in a noisy
channel. The analysis is performed by assigning a loss $\LB$ to each round,
and losses $\LA, \LU$ to false acceptance and false rejection respectively.

We show how a nearly-optimal threshold $\hthr$ for a given number of
rounds $n$ can be chosen and give {\em worst-case} bounds on the
\textit{expected loss} for that choice.  Thus, the bounds hold no
matter if the party that attempts to get authenticated is either a
legitimate user $U$ or an attacker $A$. This extends our previous work
\cite{sheddingLight}, which proposed a new \textit{distance bounding}
protocol ({\Hitomi}) and only calculated a value for the threshold
$\tau$, without providing any bounds.

We also show how a nearly-optimal number of rounds $\hns$ can be
chosen and give further bounds on the \textit{expected loss}.  The
bounds hold for {\em any bounded} error function, and not only for
$\{0, 1\}$ errors.\footnote{In all previous proposals, there is either
  an error at each round, or there is not.}  Furthermore, they are
valid for any $n$, since they are based on probability inequalities
for a finite number of samples.\footnote{The analysis
  in~\cite{singelee2} only holds for large $n$, so the approximation
  only holds asymptotically.} Thus, they are considerably more general
to the bounds of~\cite{singelee2}.

Finally, we provide high-probability estimates for the current noise
level in the channel by leveraging the coding performed in the initial
and final phases of the protocol, which take place in a coded
channel. This enables us to significantly weaken assumptions on
knowledge of the noise level in the channel and in turn, provide an
authentication algorithm which has low expected loss with high probability.






 









\section{Preliminaries}
\label{sec:preliminaries}

We consider sequences $x = x_1, \ldots, x_n$ with all $x_i$ in some
alphabet $\CX$ and $x \in \CX^n$. We write $\CX^* \defn
\bigcup_{n=0}^\infty \CX^n$ for the set of all sequences. 
We use $\defn$ to indicate a definition.  $\Pr(A)$ denotes the
probability of event $A$, while $\E$ denotes expectations so that
$\E(X \mid A) = \sum_{u \in \Omega} u \Pr(X = u \mid A)$ denotes the
conditional expectation of random variable $X \in \Omega$ when $A$ is
true. The notation $\Borel{\Omega}$ will denote an appropriate
$\sigma$-field on $\Omega$. Finally, $\ind{A}$ is an indicator
function equal to $1$ when $A$ is true and $0$ otherwise.

We consider shared secret challenge-response authentication protocols
with multi-round exchanges.  In such protocols, a verifier $\ver$
grants access to a prover $\auth$, if the latter can demonstrate its
identity $\id$ and possession of a shared secret $s \in \CX^m$. The
protocol has three phases:
\begin{inparaenum}[(i)]
\item An initialisation phase.
\item A \textit{rapid-bit exchange} phase, lasting $n$ rounds.
\item A termination phase.
\end{inparaenum}
A frequent assumption is that the authentication takes place in a
noise-free channel. The extension to noisy channels is done by
assuming the existence of an error correcting protocol. Thus, the {\em
  error} analysis is performed separately from the {\em cryptographic}
analysis. Here we shall integrate the two aspects of the problem by
performing an {\em expected loss} analysis of the authentication
protocol directly on the noisy channel. We shall assume that the
initialisation and termination phases are fixed (due to other security
considerations) and focus on the \textit{rapid-bit exchange} phase.

Due to noise in the physical medium, in any exchange between $\ver$
and $\auth$, the former may send a symbol $x \in \CX$, while the
latter may receive a symbol $\rx \in \CX$ such that $x \neq \rx$. We
shall denote the probability of erroneous transmission in the data
layer as: $\noise \defn \Pr(x \neq \rx)$, $\forall x, \rx \in \CX$.
For simplicity, we shall only treat the case of symmetric channel
noise such that: $\Pr(\rx = y \mid x \neq \rx) = \frac{1}{|\CX| - 1}$,
$\forall y \neq x, ~x,y \in \CX$.

\subsection{Thresholded protocols}
\label{sec:thresholded-protocols}
During {\em multi-round challenge response authentication} phase
(e.g. the \textit{rapid-bit exchange} phase in an RFID
distance-bounding protocol) the verifier $\ver$ sends $n$ challenges
$c_1, \ldots, c_n$, with $c_k \in \CX$, to the prover $\auth$, which
responds by transmitting $n$ responses $r_1, \ldots, r_n$, with $r_k
\in \CX$. We use $c = (c_k)_{k=1}^n$, and $r = (r_k)_{k=1}^n$ to
denote the complete challenge and response sequences respectively. The
verifier $\ver$ can calculate the correct responses $R(c_i)$ and so
can calculate an error $\err_i$ for the $i$-th round. While the
legitimate user $U$ should also be able to calculate the correct
responses, due to noise, there may be errors in the received
responses. On the other hand, the attacker has to resort to guessing,
so the expected error of the attacker should be higher than that of
the user.  In order to trade off false acceptances with false
rejections, we need a threshold value $\thr$, such that a prover is
accepted if and only if the total error observed is smaller than
$\thr$. More precisely, we define:
\begin{definition}
  An additive thresholded multi-round challenge-response
  authentication protocol has the following parameters:
  \begin{enumerate}
  \item A natural number $n > 0$, equal to the number of
    challenge-response rounds.
  \item A threshold $\thr \geq 0$.
  \item An error function $\err \defn \sum_{i=1}^n \err_{i}$, where
    $\err_{i} \in [0,1]$ represents the error of the $i$-th round.
  \end{enumerate}
  The verifier $\ver$ rejects the prover (authenticator) $\auth$,
  if and only if $\err \geq \thr$.
  \label{def:thresholded-auth}
\end{definition}
The relation of $\err$ to the challenge and response strings $c$ and
$r$ strongly depends on the protocol.  In order to make our analysis
generally applicable, we define $\pa \leq \E(\err_i \mid A)$, a lower
bound on the expected per-round error of the attacker and $\pu \geq
\E(\err_i \mid U)$, an upper bound on the error of a legitimate user.
These bounds depend on the noise and on the protocol under
consideration.  We shall return to them in
section~\ref{sec:specific-protocols}.

\section{Expected loss analysis}
\label{sec:loss-analysis}
We now specify our potential losses. For every round of the
\textit{rapid-bit exchange phase}, we suffer loss $\LB$. In addition,
we suffer a loss of $\LA$ for each false acceptance and a loss $\LU$
for false rejection.\footnote{These losses are subjectively set to
  application-dependent values. Clearly, for cases where falsely
  authenticating an attacker the impact is severe, $\LA$ must be much
  greater than $\LU$.} Given that we perform $n$ rounds, the total
loss when the prover $\auth$ is either the legitimate user $U$ or the
attacker $A$ is given by:
\begin{equation}
  \loss =
  \begin{cases}
    n\LB + \LU, & \textrm{if $\err \geq \thr$ and $\auth = U$}\\
    n\LB + \LA, & \textrm{if $\err < \thr$ and $\auth = A$}\\
    n\LB,  & \textrm{otherwise}.
  \end{cases}
  \label{eq:loss}
\end{equation}
Armed with this information, we can now embark upon an
expected loss analysis.  We wish to devise an algorithm that
guarantees an {\em upper bound} on the expected loss $\E \loss$. To
start with, we note that the expected loss when the
communicating party is an attacker $A$ or the user $U$, is given
respectively by:
\begin{align}
  \E (\loss \mid A) &= n\LB + \Pr(\err < \thr \mid A) \cdot \LA +
  \Pr(\err \geq \thr \mid A) \cdot 0
  \label{eq:attacker-loss}
  \\
  \E (\loss \mid U) &= n\LB + \Pr(\err < \thr \mid U) \cdot 0 +
  \Pr(\err \geq \thr \mid U) \cdot \LU.
  \label{eq:user-loss}
\end{align}
The \textit{expected loss} is in either case bounded by the {\em worst-case expected loss}:
\begin{equation}
  \worst \defn \max \left\{ \E(\loss \mid A), \E(\loss \mid U)\right\}
\geq \E \loss 
  \label{eq:expected-loss}
\end{equation}
If we can find an expression that bounds both $\E(\loss \mid A)$ and
$\E(\loss \mid U)$, we automatically obtain a bound on the expected
loss, $\E \loss$.

The remainder of this section is organised as
follows. Section~\ref{sec:threshold-choice} shows how a nearly-optimal
threshold $\hthr$ for a given number of rounds $n$ can be chosen and
gives bounds on the expected loss for that choice.
Section~\ref{sec:rounds-choice} shows how a nearly-optimal number of
rounds $\hns$ can be chosen and gives further bounds.
\subsection{Choice of threshold}
\label{sec:threshold-choice}
We want to choose a threshold $\tau$ such that no matter whether the
prover $\auth$ is the attacker $A$ or the legitimate user $U$ the
expected loss $\E (\loss \mid \auth)$ is as small as possible.  The
problem is that as we \textit{increase} the threshold $\tau$,
$\E(\loss \mid \auth=U)$ \textit{decreases}, while $\E(\loss \mid
\auth=A)$ \textit{increases}. The opposite is happening when we
\textit{decrease} the threshold $\tau$.  Thus, to minimise the
worst-case expected loss, we can choose a threshold $\tau$ such that
$\E(\loss \mid \auth=A, \tau) =\E(\loss \mid \auth=U, \tau)$.  A
particular choice of the threshold $\tau$ that minimises an upper
bound on the worst-case expected loss is given in
Theorem~\ref{thm:threshold}.  As an intermediate step, we obtain a
bound on the worst-case expected loss for {\em any} given threshold
$\tau$. Formally, we can show the following:
\begin{lemma}
  Let $\err_i \in [0,1]$ be the error of the $i$-th round.  If, for all
  $i > 0$, it holds that $\bar{z}_A \defn \E(\err_i \mid A) \geq \pa$
  and $\bar{z}_U \defn \E(\err_i \mid U) \leq \pu$, for some $\pa, \pu
  \in [0,1]$ such that $n\pa \leq \thr \leq n\pu$, then:
  \begin{align}
    \ELb(n;\thr)
    &\defn n\LB +
    \techreport{\max\left\{\exp\left(-\frac{2}{n}(n\pu - \thr)^2\right) \LU,
      \exp\left(-\frac{2}{n}(n\pa - \thr)^2\right)  \LA\right\}}
    \ieeetrans{\max\left\{e^{-\frac{2}{n}(n\pu - \thr)^2} \LU,
      e^{-\frac{2}{n}(n\pa - \thr)^2}  \LA\right\}}
 \nonumber
 \\
 &
    \geq \max \left\{ \E(\loss \mid A), \E(\loss \mid U)\right\}
    \geq \E \loss.     \label{eq:hoeffding-loss-bound}
  \end{align}
\end{lemma}
\begin{IEEEproof}
  The expected loss when $\auth = A$, is simply:
  \begin{align*}
    \E(\loss \mid A)
    &= n\LB + \Pr\left(\sum_i \err_i < \thr ~\Big|~ A\right) \LA
    \\
    & 
    = n\LB + \Pr\left(\sum_i \err_i - n \bar{z}_A < \thr - n \bar{z}_A ~\Big|~ A\right) \LA
    \\
    &\leq n\LB + \Pr\left(\sum_i \err_i - n \bar{z}_A < \thr - n\pa ~\Big|~ A\right) \LA
    \\
    &
    \leq n\LB + e^{-\frac{2}{n}(n\pa - \thr)^2} \LA,
  \end{align*}
  the last two steps used the fact that $\bar{z}_A \geq \pa$ and the
  \textit{Hoeffding inequality}~\eqref{eq:hoeffding}. Specifically, in
  our case, Lemma~\ref{lem:hoeffding} (page~\pageref{lem:hoeffding})
  applies with $X_i = \epsilon_i$. Then, it is easy to see that $\mu_i
  = \bar{z}_A$ for all $i$ and $b_i - a_i = 1$, so $\Pr( \epsilon < n
  \pa + n t \mid A) \leq e^{-2nt^2}$.  By setting $\thr = n\pa + nt$,
  we obtain $t = (\thr - n\pa)/n$, which we can plug into the above
  inequality, thus arriving at the required result.  The user case,
  $\auth = U$, is handled similarly and we conclude that $\E(\loss
  \mid U) \leq n\LB + e^{-\frac{2}{n}(n\pu - \thr)^2} \LU$.
\end{IEEEproof}
Having bounded the loss suffered when choosing a specific threshold,
we now choose a threshold $\hthr$ that minimises the above bound for
fixed $n$. In fact, we can show that such a threshold results in a
particular loss bound.
\begin{theorem}
  Let $\rho \defn \LA / \LU$ and select
  \begin{equation}
    \thr = \hthr \defn
\frac {n(\pa + \pu)} {2} - \frac {\ln{\rho}} {4\Delta}
    \label{eq:opt-threshold}
  \end{equation}
  If $n\pu \leq \tau \leq n \pa$, then the expected loss $\E \loss$ is
  bounded by:
  \begin{equation}
    \E (\loss \mid n, \hthr) \leq \ELb_1(n) \defn n\LB + e^{-\frac{n}{2}\Delta^2} \cdot \sqrt{\LA \LU}.
    \label{eq:threshold-loss-bound}
  \end{equation}
  with $\Delta \defn \pa - \pu$.
  \label{thm:threshold}
\end{theorem}
\begin{IEEEproof}
  Substitute \eqref{eq:opt-threshold} in
  the first exponential of \eqref{eq:hoeffding-loss-bound} to obtain:
  \begin{align*}
    e^{-\frac{2}{n}(n\pu - \hat{\tau}^*)^2} \LU
\techreport{
  &=
  \exp\left(
    -\frac{n}{2}\Delta^2 + \frac{1}{2} \ln \rho - \frac{\ln^2\rho}{8n\Delta^2} + \ln \LU
  \right) \LU
  \\}
&=
e^{-\frac{n}{2}\Delta^2} e^{-\frac{\ln^2\rho}{8n\Delta^2}} \sqrt{\LA\LU}.
  \end{align*}
  It is easy to see that the exact same result is obtained by
  substituting \eqref{eq:opt-threshold} in the second exponential of \eqref{eq:hoeffding-loss-bound}.
Thus, both
  $\E(\loss \mid A)$ and $\E(\loss \mid U)$ are bounded by the same
  quantity and consequently, so is $\max\left\{\E(\loss \mid A),
    \E(\loss\mid U)\right\}$. Thus,
  \begin{align*}
    \ELb(n, \hthr)
    &\leq
    \techreport{n\LB + \exp\left(-\frac{n}{2}\Delta^2\right) \cdot \exp\left(-\frac{\ln^2\rho}{8n\Delta^2}\right) \cdot \sqrt{\LA \LU}
    \\
    &\leq}
    n\LB + 
    e^{-\frac{n}{2}\Delta^2} \sqrt{\LA \LU},
  \end{align*}
  where we simplified the bound by noting that
  $\frac{\ln^2\rho}{8n\Delta^2} > 0$.
\end{IEEEproof}
The intuition behind the algorithm and the analysis is that it is
possible to bound the probability that $A$ makes less errors than
expected, or that $U$ makes more than expected.  For this reason, the
$\hthr$ chosen in the theorem must lie between $n\pu$ and $n\pa$. This
also implies a lower bound on the number of rounds $n$.  

\subsection{Choice of the number of rounds}
\label{sec:rounds-choice}
Using similar techniques to those employed for obtaining a suitable
value for the threshold, we now indicate a good choice for the number
of rounds $n$ and provide a matching bound on the expected loss.
\begin{theorem}
  Assume $\LA, \LU, \LB > 0$. If we choose $\thr = \hthr$
  and
  \begin{equation}
    n = \hns \defn \frac{\sqrt{1 + 2CK} - 1}{C},
    \label{eq:opt-rounds}
  \end{equation}
  where $C = \Delta^2$ and $K = \sqrt{\LA\LU}/\LB$, then the expected
  loss $\E \loss$ is bounded by:
  \begin{equation}
    \E (\loss \mid \hthr, \hns) \leq \ELb_2 \defn \sqrt{8K/C} \cdot \LB
    =
    \frac{\sqrt{8 \LB} (\LA\LU)^{1/4}}{\Delta}.
    \label{eq:loss-opt-rounds}
  \end{equation}
  \label{the:rounds}
\end{theorem}
\begin{IEEEproof}
  We shall bound each one of the summands of \eqref{eq:threshold-loss-bound}
  by $\sqrt{2K/C}\cdot \LB$.
  For the first term we have:
  \begin{align*}
    n\LB &= \frac{\sqrt{1 + 2CK} - 1}{C} \LB
    \leq
    \frac{\sqrt{1 + 2CK}}{C} \LB
    \\
    &\leq
    \frac{\sqrt{2CK}}{C} \LB
    =
    \sqrt{\frac{2K}{C}} \LB.
  \end{align*}
  For the second term, by noting that $e^x \geq 1 + x$, we have:
  \begin{align*}
     \sqrt{\LU\LA} 
     \cdot e^{-\frac{n}{2}\Delta^2}
     &\leq
     \frac{\sqrt{\LU\LA}}
     {1+\frac{n}{2}\Delta^2}
     =
     \frac{K\LB}
     {1+\frac{nC}{2}}
     \techreport{=
       \frac{2K\LB}
     =
     \frac{2K\LB}
     {2 + nC}
   }
     =
     \frac{2K\LB}
     {1 + \sqrt{1 + 2CK}}
     \\
     &\leq
     \frac{2K\LB}
     {\sqrt{1 + 2CK}}
     \leq
     \frac{2K\LB}
     {\sqrt{2CK}}
     =
     \sqrt{\frac{2K}{C}} \LB.
  \end{align*}
  Summing the two bounds, we obtain the required result.
\end{IEEEproof}

This theorem proves that our {\em worst-case expected loss} $\worst$
grows sublinearly both with increasing round cost (with rate
$O(\epsilon^{1/2})$) and with increasing authentication costs (with
rate $O(\epsilon^{1/4}))$. Furthermore, the {\em expected loss} is
bounded symmetrically for both user and attacker access. Finally,
there is a strong dependence on the margin $\Delta$ between the
attacker and the user error rates, which is an expected result.

\section{Analysis of RFID thresholded protocols}
\label{sec:specific-protocols}

Currently, the only known protocols employing an authentication
phase without any error correction that we are aware of are {\em RFID
  distance bounding} protocols. For that reason, we shall examine the
properties of two such protocols, for which it is possible to derive
expressions for $\pa, \pu$ given a symmetric channel noise $\omega$.

The {\Swiss} protocol~\cite{KimAKSP-2008-icisc} 
and the
variant {\Hitomi}~\cite{sheddingLight}
are thresholded authentication protocols satisfying
Definition~\ref{def:thresholded-auth}.  It is easy to show (for
details see~\cite{sheddingLight}) that, for those two protocols, under
channel noise $\noise$, the expected error bounds $\pa, \pu$ are given
by: $\pa = \frac {\noise+1} {2}$, $\pu =2 \noise$, where we note in
passing that $\pa \geq \pu$ and so $\noise \leq \frac {1} {3}$.
Finally, by substituting $\Delta = \pa - \pu= \frac{1-3 \noise} {2}$
in \eqref{eq:threshold-loss-bound}, we obtain the following bound for
{\Swiss} and {\Hitomi}:
\begin{equation}
  \E \loss \leq n\LB + e^{-\frac{n(1-3\noise)^{2}}{8}} \cdot \sqrt{\LA\LU},
  \label{eq:swiss-knife-threshold}
\end{equation}

We have performed a number of experiments to test the efficacy of
these protocols, when used in conjunction with our suggested, as well
as the optimal values of the threshold and number of rounds.  In all
of the experiments shown here, we chose the following values for the
losses: $\LA = 10$, $\LU = 1$, $\LB = 10^{-2}$.  

\begin{figure}
\centering
\subfigure[The \textit{Expected Loss} $\E L$ and the bound on the \textit{Expected Loss} $\ELb_1$ vs. the number of bits $n$ exchanged during the rapid single bit exchange for various values of channel noise $\noise$.]
{
  \label{fig:ExpectedLossVsN-a}
  \psfrag{T1, BER=0.01}[r][r][0.75][0]{$\ELb_1(n)$, $\noise=10^{-2}$}
  \psfrag{EL, BER=0.01}[r][r][0.75][0]{$\max_\auth \E(\loss \mid \auth, n)$, $\noise=10^{-2}$}
  \psfrag{T1, BER=0.1}[r][r][0.75][0]{$\ELb_1(n)$, $\noise=10^{-1}$}
  \psfrag{EL, BER=0.1}[r][r][0.75][0]{$\max_\auth \E(\loss \mid \auth, n)$, $\noise=10^{-1}$}
  \psfrag{hn number of bits}[B][B][0.75][0]{$n$ (number of rounds)}
  \psfrag{Expected Loss}[B][B][0.75][0]{Expected loss}
  \includegraphics[width=0.95\columnwidth]{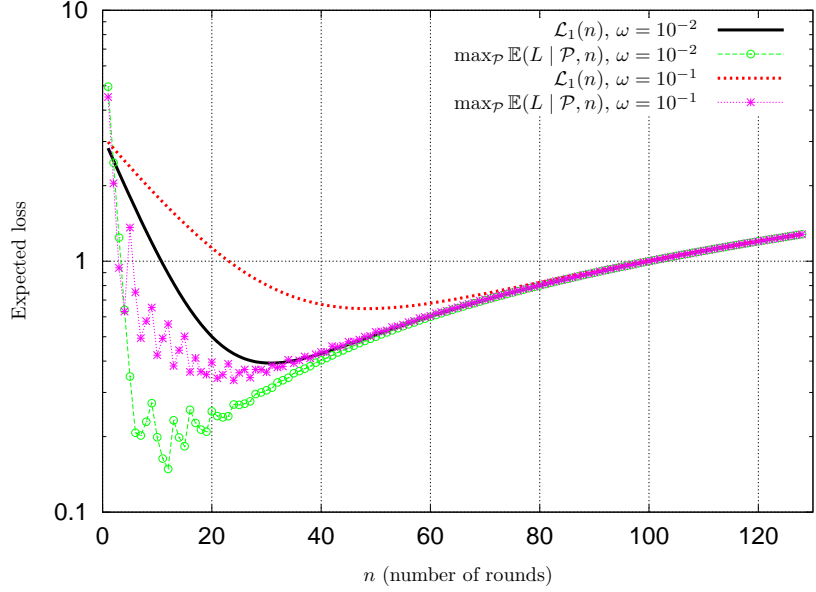}
}
\subfigure[The worst-case expected Loss $\worst$ and the bounds $\ELb_1$ and
$\ELb_2$ from theorems 1 and 2 respectively vs. the channel error rate
$\noise$.]{
  \label{fig:expected-losses-noise-nopt}
  \psfrag{ns}[r][r][0.75][0]{$\max_\auth \E(\loss \mid \auth, \ns)$}
  \psfrag{hns}[r][r][0.75][0]{$\max_\auth \E(\loss \mid \auth, \hns)$}
  \psfrag{ELBT1}[r][r][0.75][0]{$\ELb_1(\hns)$}
  \psfrag{ELBT2}[r][r][0.75][0]{$\ELb_2$}
  \psfrag{bit error rate}[B][B][0.75][0]{$\noise$ (channel noise)}
  \psfrag{Expected Loss}[B][B][0.75][0]{Expected loss}
  \includegraphics[width=0.95\columnwidth]{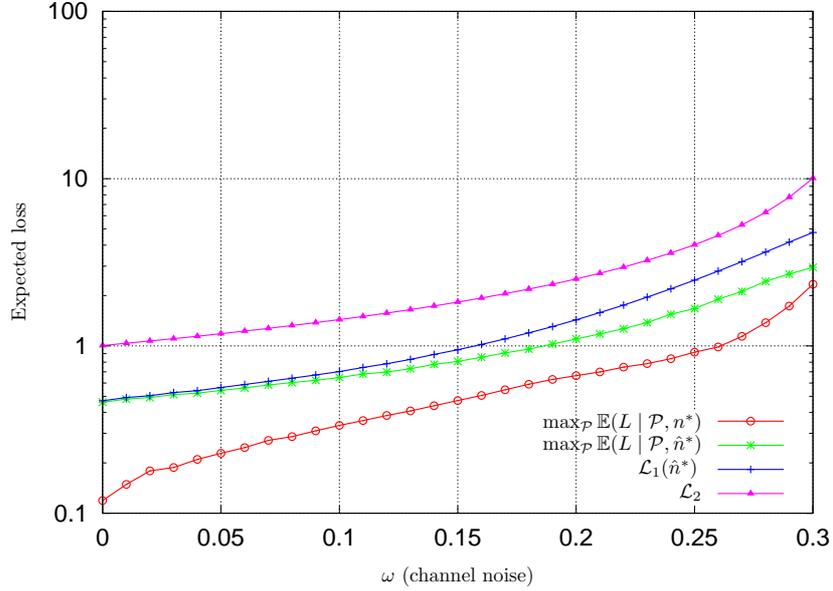}
}
\caption{Comparison of all losses.}
\label{fig:losses-rounds}
\end{figure}

Figure~\ref{fig:ExpectedLossVsN-a} depicts the bound
\eqref{eq:swiss-knife-threshold} on the expected loss, as well as the
actual $\E \loss$ calculated via the binomial formula, when the
threshold $\hthr$, calculated from (\ref{eq:opt-threshold}), is used.
We plot both the expected loss and the bound for two different channel
noise levels $\noise \in \set{10^{-1}, 10^{-2}}$, where the number of
rounds $n$ varies from $1$ to $256$. Obviously, the bound is greater
than the actual expected loss, while it approaches it exponentially
fast as $n$ increases.  In addition, the losses are higher when the
amount of noise increases.

Furthermore, we can see that there are minimising values of $n$ for
all cases. While they do not coincide for the bound and the actual
\textit{expected loss}, they are within a factor of two of each
other. Finally, $\hns$, the value of $n$ minimising the bound, is
always greater than $\ns \defn \argmin_n \max_{\auth \in A, U} \E
(\loss \mid \auth, n)$, the value of $n$ that minimises the {\em
  worst-case expected loss}.  Since the probability of incorrect
authentication always decreases with increasing $n$, this implies that
any additional losses incurred by using $\hns$ is due to transmission
costs only.

Figure~\ref{fig:expected-losses-noise-nopt}
examines the effect of noise in more detail.  In particular,
it depicts the \textit{worst-case expected loss} for the optimal number of
rounds $\ns$, denoted by $\E (\loss \mid \ns)$ in the legend.  This is
of course smaller than $\E (\loss \mid \hns)$, the loss suffered by
choosing $\hns$, with the gap becoming smaller for larger error
rates. Since when this occurs, the \textit{expected loss} is very close to
$\ELb_1$, this implies that the bound of the Theorem~1 needs considerable
tightening for small $\noise$. Finally, $\ELb_2$ is considerably
looser, and thus it is only of theoretical interest.

Finally, due to the way that the protocols under consideration
generate challenges and responses, the number of rounds $n$ must be
smaller than the length $k$ of the messages in the initialization
phase and also the length of the key $x$.  Thus, in practice we will
always select a number of rounds $n = \min \{ n^*, k\}$.  This
condition is necessary since the responses $r$ for both protocols are
calculated using an XOR operation between the secret key $x$ and a
constant value (i.e. either $\alpha$ or $\beta$) that has the same
length $k$. Since these protocols are only used as examples, it is
beyond the scope of this paper to propose protocols that do not suffer
from this limitation.

\section{Estimating $\noise$}
\label{sec:noise}
In this section, we discuss how it is possible to calculate the
channel error rate $\noise$, which is used in the expressions for
$\pa, \pu$. This can be done by leveraging the coding performed during
the initial and final phases of the protocol.  We assume some coding
function $\code : \CX^m \to \CX^k$, with $k > m$, and a metric
$\metric$ on $\CX^k$ (where usually $\CX=\{0,1\}$ and $\metric$ is the
Hamming distance) such that:
\begin{equation}
\minmetric \defn 
\cset{\metric(\code(x), \code(y))}{x,y \in \CX^m, x \neq y}
\label{eq:min-metric}
\end{equation}
is the minimum (Hamming) distance between valid codewords.  For a
given $x \in \CX^m$, the source transmits $\phi = \code(x)$ and the
sink receives $\hat{\phi}$, with $\phi, \hat{\phi} \in \CX^n$.  As
before, we assume that the physical channel has a symmetric error rate
$\noise = \Pr(\hat{\phi_i} \neq \phi_i)$, where $\phi_i$ denotes the
$i$-th bit of $\phi$.  This is then decoded as $\hat{x} \defn
\argmin\cset{\metric(\hat{\phi}, \code(y))}{y \in \CX^m}$. Let $\theta$ be
the number of errors in the string $\hat{\phi}$, or more precisely $\theta
= \metric(\phi, \hat{\phi})$.  Let $\hat{\theta} \defn
\metric(\code(\hat{x}), \hat{\phi})$ be the distance between the
closest valid codeword $\code(\hat{x})$ and the received
$\hat{\phi}$. If $\theta < (\minmetric - 1)/2$, then $\theta = \hat{\theta}$.


The crux of our method for estimating $\omega$ relies on the number of
errors $\theta$ being less than $(\minmetric - 1)/2$, in which case, the
estimated number of errors $\hat{\theta}$ will equal $\theta$.  Let
$\hat{\noise} \defn \frac{\hat{\theta}}{n}$ be our empirical error rate.
In that case, the expected empirical error rate equals the true error
rate. More formally:
\begin{equation}
  \label{eq:erorr-rate-estimate}
  \E\left(\hat{\noise} \mid
    \theta \leq (\minmetric - 1)/2 \right)
  =
  \noise.
\end{equation}
If $\theta > (\minmetric - 1)/2$ then the protocol fails in any case, due
to decoding errors in the initial or final phases.  If not, then the
above equation holds and we can obtain high probability bounds for
$\omega$ via the \textit{Hoeffding inequality} (Appendix,  Lemma \ref{lem:hoeffding}).  In particular, it is easy to
show that, for any $\delta \in [0, 1]$:
\begin{equation}
  \label{eq:high-probability-bound}
  \Pr\left(|\hat{\noise} - \noise| \geq \sqrt{\frac{\ln2/\delta}{2k}}\right)
  \leq \delta,
\end{equation}
by substituting the square-root term into (\ref{eq:hoeffding}), and
setting $\mu_i = \noise$, $\sum X_i = \hat{\theta}$, $a_i = 0$, $b_i=1$.
Consequently, for the {\Swiss} family of protocols the following
values for $\pa$ and $\pu$ hold with probability $1 - \delta$:
\begin{align}
  \label{eq:pa-hp}
  \pa &= \frac{1 + \hat{\noise}}{2}
  + \sqrt{\frac{\ln2/\delta}{8k}},
  &
  \pu &= 2\hat{\noise}
  - \sqrt{\frac{2\ln2/\delta}{k}}.
\end{align}
While we were unable to provide bounds on the performance of this
choice, experimental investigations presented in the next section
indicate that it has good performance.

\subsection{Evaluation Experiments}
We have performed some experiments to evaluate our methods in a more
realistic setting, involving an RFID distance bounding protocol with a
\textit{rapid-bit exchange} phase.  We perform simulations for two
cases: Firstly, when a \textit{legitimate user} $U$ is trying to get
authenticated and secondly, when an \textit{adversary} $A$ is trying
to perform a mafia fraud attack.  We have estimated the
\textit{worst-case expected loss} by running $10^4$ experiments for
each case, obtaining a pair of estimates $\hat{\E}(\loss \mid A)$,
$\hat{\E}(\loss \mid U)$ by averaging the loss $L$, as defined in
\eqref{eq:loss}, incurred in each experiment and taking the maximum of
the two.  In all of the experiments shown in this section, we chose
the following values for the losses: $\LA = 10$, $\LU = 1$, $\LB =
10^{-2}$, while we used $k=2^{10}$ for the coded messages in the
initialisation phase.

The actual values $\pa, \pu$ depend on $\noise$, which is unknown. We
compare three methods for choosing $\pa, \pu$. Firstly, guessing a
value $\hat{\noise}$ for the channel noise. Secondly, using the maximum
likelihood noise estimate $\hat{\noise} = \hat{\theta} / k$.  In both
cases, we simply use $\hat{\omega}$ as described at the beginning of
Sec.~\ref{sec:specific-protocols} to obtain $\pa, \pu$.  In the third
case, we use the high-probability bounds~\eqref{eq:pa-hp} for $\pa,
\pu$, with an arbitrary value of $\delta$.

In the first experiment, we use the nearly-optimal threshold and
number of rounds that we have derived in our analysis.  In the second
experiment, we replace our choice of threshold with a choice similar
to that of Baign\`{e}res et al.~\cite{ProvSec2010}. Their threshold is
derived via the likelihood ratio test, which is asymptotically optimal
(c.f. ~\cite{Degroot:OptimalStatisticalDecisions,Chernoff:SequentialDesignExperiments})
\begin{equation}
\acute{\tau}= \frac {n \ln \frac {1-\pu} {1-\pa}}  {
\ln \frac {1-\pu} {1-\pa} - {\ln{ \frac {\pu}{\pa}}}}
\end{equation}
Since in our case we have unequal losses $\ell_A$ and $\ell_U$, we
re-derive their threshold via a Bayesian test (to which a Bayesian
formulation of the Neymann-Pearson
lemma~\cite{Degroot:OptimalStatisticalDecisions} applies) to obtain:
\begin{equation}
\tilde{\tau}= \frac {n \ln \frac {1-\pu} {1-\pa} - \ln \rho}  {
\ln \frac {1-\pu} {1-\pa} - {\ln{ \frac {\pu}{\pa}}}}
\label{eq:asymptotic-threshold}
\end{equation}
Interestingly, for small $\Delta$, the form of $\tilde{\tau}$ is
similar to $\hthr$: Let $\bar{p}$ such that $\pa = \bar{p} + \Delta/2$
and $\pu = \bar{p} - \Delta/2$. Then (\ref{eq:asymptotic-threshold})
can be approximated by:
\begin{equation}
\tilde{\tau^*}=    n\bar{p} - \frac {  \bar{p}(1-\bar{p})   } {\Delta} \ln{\rho}
\end{equation}
More details on the derivation of \eqref{eq:asymptotic-threshold} are
given in Appendix~\ref{app:asymptotics}.

Figure~\ref{fig:loss-comparison} depicts the \textit{worst-case
  expected loss} $\worst$ as a function of the actual noise $\noise$.
Figure~\ref{fig:LossComparison-a} shows $\worst$ using the threshold
$\tau$ derived from our {\em expected loss}
analysis~\eqref{eq:opt-threshold}, while in
Figure~\ref{fig:LossComparison-b} we use the asymptotically optimal
threshold of \eqref{eq:asymptotic-threshold}. In both cases, we plot
$\worst$, while the actual noise $\noise$ is changing, for a number of
different cases. Initially, we investigate the evolution of $\worst$
for three arbitrarily chosen values $\hat{\noise} \in \{10^{-1},
10^{-2}, 10^{-3}\}$. Additionally, we examine the evolution of the
{\em worst-case expected loss}, when the noise is empirically
estimated $\hat{\noise} = \hat{\theta} / n$ and finally when $\pa$ and
$\pu$ are calculated via equation (\ref{eq:pa-hp}) with $\delta \in \{
10^{-1}, 10^{-2}\}$.


As it can be seen in Figure~\ref{fig:loss-comparison}, in all cases
(using ours Figure~\ref{fig:LossComparison-a} or Baign\`{e}res et
al. \cite{ProvSec2010} threshold Figure~\ref{fig:LossComparison-b})
the \textit{worst-case expected loss} is very low for small values of
the actual noise and increases sharply when the actual noise exceeds
the value of $10^{-1}$. It is interesting to see that when we use the
optimistic\footnote{Experiments with pessimistic high probability
  estimates for the noise showed a significant increase in the number
  of rounds used, which resulted in a higher expected loss.}
high probability estimates for $\pa, \pu$, we obtain almost always
better performance than simply guessing the noise, or using the plain
empirical estimate $\hat{\noise}$ directly. Furthermore, using the
asymptotically optimal threshold \eqref{eq:asymptotic-threshold}, we
observe a deterioration in the results.


As mentioned in the related work (Sec.~\ref{sec:relatedwork}), the
choice of the threshold by Baign\`{e}res et al. \cite{ProvSec2010} is
only asymptotically optimal. Ours, while not optimal, gives a
worst-case expected loss guarantee for any finite sample size. Thus,
it has better performance when the asymptotic approximation is not
sufficiently good, which occurs when both the number of rounds $n$ and
the gap $\Delta$ are small.

\begin{figure}
\centering
\subfigure[Our threshold]{
 \label{fig:LossComparison-a}
 \psfrag{omega}[B][B][1][0]{$\noise$}
 \psfrag{Expected Loss}[B][B][0.75][0]{Expected loss}
 \psfrag{hw=0.1}[B][B][0.55][0]{$\hat{\noise}=10^{-1}$}
 \psfrag{hw=0.01}[B][B][0.55][0]{$\hat{\noise}=10^{-2}$}
 \psfrag{hw=0.001}[B][B][0.55][0]{$\hat{\noise}=10^{-3}$}
 \psfrag{hw=k/n}[B][B][0.55][0]{$\hat{\noise}=\theta/k$}
 \psfrag{d=0.1}[B][B][0.55][0]{$\delta=10^{-1}$}
 \psfrag{d=0.01}[B][B][0.55][0]{$\delta=10^{-2}$}
 \includegraphics[width=0.95\columnwidth]{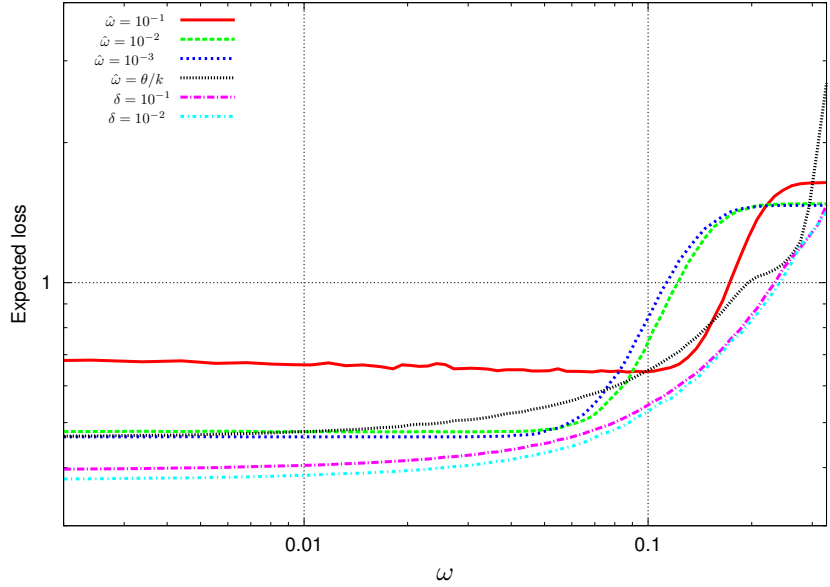}
}
\subfigure[Asymptotic threshold]{
 \label{fig:LossComparison-b}
 \psfrag{omega}[B][B][1][0]{$\noise$}
 \psfrag{Expected Loss}[B][B][0.75][0]{Expected loss}
 \psfrag{hw=0.1}[B][B][0.55][0]{$\hat{\noise}=10^{-1}$}
 \psfrag{hw=0.01}[B][B][0.55][0]{$\hat{\noise}=10^{-2}$}
 \psfrag{hw=0.001}[B][B][0.55][0]{$\hat{\noise}=10^{-3}$}
 \psfrag{hw=k/n}[B][B][0.55][0]{$\hat{\noise}=\theta/k$}
 \psfrag{d=0.1}[B][B][0.55][0]{$\delta=10^{-1}$}
 \psfrag{d=0.01}[B][B][0.55][0]{$\delta=10^{-2}$}
 \includegraphics[width=0.95\columnwidth]{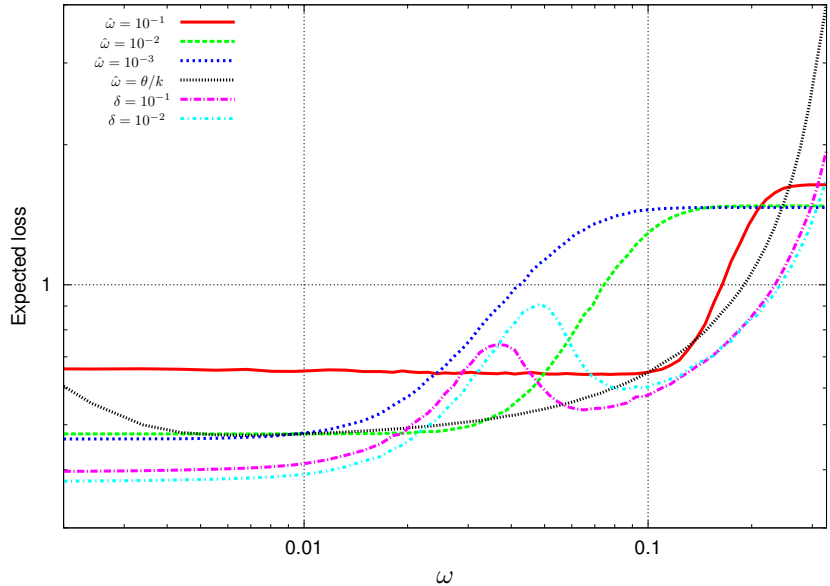}
}
\caption{The worst-case expected loss as a function of noise. We plot
 the evolution of the loss as noise changes, for a number of
 different cases. Firstly, for the case where we arbitrarily assume a
 noise value $\hat{\noise} \in \{10^{-1}, 10^{-2},
 10^{-3}\}$. Secondly, for an empirically estimated $\hat{\noise} =
 \hat{k} / n$, and finally for $\pa, \pu$ calculated via equation
(\ref{eq:pa-hp}) with $\delta \in \{ 10^{-1}, 10^{-2}\}$.}
\label{fig:loss-comparison}
\end{figure}

Finally, note that in the {\Swiss} and {\Hitomi} protocols, the size
of the initialisation messages is fixed, since the number of rounds
$n$ is fixed.  In practice, one would have to modify these protocols
in order for them to work with an arbitrary number of rounds, but this
subject is beyond the scope of this paper. Our focus is mainly the
{\em expected loss} analysis of the noisy authentication phase.

\section{Conclusion}
\label{sec:conclusion} 

We have performed an expected loss analysis for thresholded
authentication protocols under noise.  This is particularly
significant for areas of communications where challenges and responses
are costly and where there exists significant uncertainty about the
correctness of any single response. More precisely, we have shown how
to select a threshold and provided an upper bound on the worst-case
expected loss. Additionally, we have shown how to select the number of
rounds in order to tighten the loss and obtained a loss bound that
holds uniformly and depends only on the error rates of the user and
attacker and the individual losses.  We have applied these choices of
threshold and number of rounds to two representative distance bounding
protocols, the {\Swiss} and the \Hitomi. In addition, we have
presented a high-probability method for estimating the channel noise
for such protocols.  We have examined its performance in further
simulation experiments with {\em unknown channel noise}, and shown
that we obtain uniformly superior results to guessing or direct
empirical noise estimates. Finally, we repeated those experiments with
a asymptotically optimal threshold similar to that used by
Baign\`{e}res et al.\cite{ProvSec2010}. Our results indicate a
significant improvement through the use of a threshold with uniform,
rather than asymptotic, guarantees. Consequently, it is our view that
algorithms motivated by an asymptotic analysis should be avoided in
the finite-sample regime of most challenge-response authentication
protocols.



\section*{Acknowledgements}
This work was partially supported by the Marie Curie IEF project
``PPIDR: Privacy-Preserving Intrusion Detection and Response in
Wireless Communications'', grant number: 252323 and by the IM-CLeVeR
EU FP7 Integrated Project, grant number: 231722.

\appendix
\techreport{\section{Auxilliary results}}
\subsection{Useful formulas}
\label{sec:binomial}
If $X_1, \ldots, X_n$ are Bernoulli random variables with $X_k \in
\set{0, 1}$ and $\Pr(X_k = 1) = \mu$ for all $k$, then
\begin{equation}
  \label{eq:binomial}
  \Pr\left(\sum_{k=1}^n X_k \geq u\right)
    =
    \sum_{k=0}^{u} \binom{n}{k} \mu^k (1 - \mu)^{n-k}.
\end{equation}
This probability can be bounded via \textit{Hoeffding's inequality} \cite{Hoeffding}:
\label{sec:Hoeffding}
\begin{lemma}[Hoeffding]
  For independent random variables $X_1, \ldots, X_n$ such that
  $X_i \in [a_i, b_i]$, with  $\mu_i \defn \E X_i$ and $t>0$:
  \begin{align}
    \Pr\left(
      \sum_{i=1}^n X_i \geq \sum_{i=1}^n \mu_i + n t
    \right) 
    \ieeetrans{&}
    = \Pr\left(
      \sum_{i=1}^n X_i \leq \sum_{i=1}^n \mu_i - n t
    \right) 
    \ieeetrans{\nonumber\\&}
    \leq
    \exp\left(
      -\frac{2n^2t^2}{\sum_{i=1}^n (b_i - a_i)^2}
    \right).
    \label{eq:hoeffding}
  \end{align}
  \label{lem:hoeffding}
\end{lemma}
\subsection{On asymptotic thresholds}
\label{app:asymptotics}
One way to obtain an asymptotically optimal threshold is to employ a
Bayesian hypothesis test~\cite{Degroot:OptimalStatisticalDecisions}.
This requires defining a prior probabilty on the possible
hypotheses. In our case, the hypothesis set is $H = \{A, U\}$, on
which we define a prior probability $\pi$. For $\{0,1\}$ errors, the
probability of observing $\err$ errors out of $n$ observations is
given by $\Pr(\err \mid A)$ and $\Pr(\err \mid U)$ for the attacker
and user respectively and it follows a binomial distribution with
parameters $\pa, \pu$ respectively. Given an observed error $x$, the
{\em posterior} probability of any hypothesis $h \in H$ is:
\[
\pi(h \mid \err = x) = \frac{\Pr(\err = x \mid h) \pi(h)}{\sum_{h' \in
    H}\Pr(\err = x \mid h') \pi(h')}.
\]
We then define a decision set $\Decs = \{\decs_A, \decs_U,
\decs_\emptyset\}$, where $\decs_A$ means we decide that the prover is
an attacker and $\decs_U$ means we decide that the prover is a user
and $\decs_\emptyset$ means that we are undecided.  Finally, we define
a loss function $L : \Decs \times H \to \Reals$, such that $L(\decs,
h)$ is our loss when we decide $\decs$ and $h$ is the correct
hypothesis.  The expected loss of decision $\decs \in \Decs$, under
our prior and given $\err$ errors out of $n$ is:
\[
\E_\pi(L \mid \err, \decs) = \sum_{h \in H} L(\decs, h) \pi(h \mid \err),
\]
where $\E_\pi$ denotes expectation with respect to the prior $\pi$.
Now define the decision function $\decfun : \{0, 1, \ldots, n\} \to
\Decs$:
\begin{equation}
\decfun(\err) \defn
\begin{cases}
  \decs_U, &\textrm{if $\E_\pi(L \mid \err, \decs_U) \leq \E_\pi(L \mid \err, \decs_A)$}
    \\
  \decs_A, &\textrm{if $\E_\pi(L \mid \err, \decs_U) > \E_\pi(L \mid \err, \decs_A)$}.
\end{cases}
\label{eq:bayes-decision}
\end{equation}
This decision function minimises $\E_\pi L$ by construction
(c.f. \cite{Degroot:OptimalStatisticalDecisions} ch. 8). 
The following
remark is applicable in our case:
\begin{remark} Assume i.i.d errors with $\err_i \in \{0, 1\}$, so that
  we can use a binomial probability for $\Pr(\err \mid h)$.  Set the
  loss function $L$ to be $L(\decs_U, A) = \LA$, $L(\decs_A, U) = \LU$
  and $0$ otherwise.  Then the decision
  function~\eqref{eq:bayes-decision} becomes equivalent to:
\[
\decfun(\err) \defn
\begin{cases}
  \decs_U, &\textrm{if $\err < \thr_b$}\\
  \decs_A, &\textrm{if $\err \geq \thr_b$},
\end{cases}
\]
where
\[
\thr_b 
\defn 
\frac {n \ln \frac {1-\pu} {1-\pa} - \ln [\rho \frac{\pi(A)}{\pi(U)}]}
{\ln \frac {1-\pu} {1-\pa} - {\ln{ \frac {\pu}{\pa}}}}
\]
\end{remark}
\begin{IEEEproof}
  We start by calculating the expected loss for either decision. First:
  \begin{align*}
    \E_\pi(L \mid \err, \decs_A) &= \LU \pi(U \mid \err) = \frac{\LU
      \pi(U) \Pr(\err \mid U)} {\pi(A) \Pr(\err \mid A) + \pi(U)\Pr(\err \mid U)},
  \end{align*}
  due to our choice of $L$ and $\pi$.  Similarly,
  \begin{align*}
    \E_\pi(L \mid \err, \decs_U) &= \LA \pi(A \mid \err) = \frac{\LA
      \pi(A) \Pr(\err \mid A)} {\pi(A) \Pr(\err \mid A) + \pi(U)\Pr(\err \mid U)}.
  \end{align*}
  Combining the above expressions, the decision
  function~\eqref{eq:bayes-decision} can then be written so that we
  make decision $\decs_U$ if and only if:
  \[
  \LA \pi(A) \Pr(\err \mid A) \leq \LU \pi(U) \Pr(\err \mid U).
  \]
  Finally, replacing \eqref{eq:binomial} with means $\pa, \pu$
  respectively and taking logarithms we obtain:
  \begin{align*}
    \ln [\rho \pi(A) / \pi(U)]  + \err \ln \frac{\pa}{\pu}
    \leq
    (n - \err) \ln \frac{1 - \pu}{1 -\pa},
  \end{align*}
  as a condition for deciding $\decs_U$. With some elementary
  manipulations, we arrive at the required result.
\end{IEEEproof}
Given
the conditions of the previous remark, it is easy to see
(c.f. ~\cite{Degroot:OptimalStatisticalDecisions} ch.~8) that the
decision function $q$ minimises the Bayes risk:
\begin{equation}
  \label{eq:bayes-risk}
  \E_\pi(L \mid q)
  =
  \pi(A) \Pr(\err < \thr_b \mid A) \LU
  + \pi(U) \Pr(\err \geq \tau_b \mid U) \LA.
\end{equation}
Furthermore, for $\pi(A) = \pi(U) = 1 / 2$, we obtain
\eqref{eq:asymptotic-threshold}.  In addition, this choice also
minimises an upper bound on the worst-case expected loss since:
\[
\max_{h \in H} \E(L \mid h, q) \leq \sum_{h \in H} \E(L \mid h, q) = 2 \E_\pi(L \mid q).
\]
for uniform $\pi$.

Finally, the asymptotic optimality of Bayesian testing generally
follows from Bayesian {\em consistency}
(c.f. \cite{Degroot:OptimalStatisticalDecisions} ch. 10). More
specifically, \cite{Chernoff:AsymptoticEfficiency} has proved the
asymptotic optimality of Bayes solutions for hypothesis testing of the
type examined here.  

\bibliographystyle{plain}
\bibliography{threshold}

\end{document}